\documentclass[%
superscriptaddress,
preprint,
 amsmath,amssymb,
 aps,
pre
]{revtex4-1}

\usepackage{graphicx}
\usepackage{dcolumn}
\usepackage{bm}
\usepackage{hyperref}

\allowdisplaybreaks 
\begin{document}


\title{Stability of anti-bunched buses and local unidirectional Kuramoto oscillators}
\author{Lock Yue Chew}
\email{lockyue@ntu.edu.sg}
\affiliation{Division of Physics and Applied Physics, School of Physical and Mathematical Sciences, 21 Nanyang Link, Nanyang Technological University, Singapore 637371}
\affiliation{Data Science and Artificial Intelligence Research Centre, Block N4 \#02a-32, Nanyang Avenue, Nanyang Technological University, Singapore 639798}
\affiliation{Complexity Institute, 61 Nanyang Drive, Nanyang Technological University, Singapore 637335}
\author{Vee-Liem Saw}
\email{Vee-Liem@ntu.edu.sg}
\affiliation{Division of Physics and Applied Physics, School of Physical and Mathematical Sciences, 21 Nanyang Link, Nanyang Technological University, Singapore 637371}
\affiliation{Data Science and Artificial Intelligence Research Centre, Block N4 \#02a-32, Nanyang Avenue, Nanyang Technological University, Singapore 639798}
\author{Yi En Ian Pang}
\affiliation{Division of Physics and Applied Physics, School of Physical and Mathematical Sciences, 21 Nanyang Link, Nanyang Technological University, Singapore 637371}
%

\date{\today}

\begin{abstract}
Inspired by our recent work that relates bus bunching as a phenomenon of synchronisation of phase oscillators, we construct a model of Kuramoto oscillators that follows an analogous interaction mechanism of local unidirectional coupling. In the bus loop system, we can introduce a no-boarding policy as a form of kicking force to achieve a stable staggered (anti-bunched) state. For Kuramoto oscillators, it turns out that such stable anti-bunched states can exist (without any additional kicking force) if the number of oscillators are at least five. This correspondence between the bus loop system and the local unidirectional Kuramoto oscillators leads to the insight on how the bus loop system can remain staggered.
\end{abstract}


\maketitle


\section{Introduction}

In a bus loop service, $N$ buses serve $M$ (reasonably) staggered bus stops around the loop, going round and round picking up and alighting passengers at the bus stops. This loop can be isometrically mapped to the unit circle, and the positions of the buses on the loop are identifiable by their phases $\theta_i\in[0,2\pi)$, $i=1,\cdots,N$ on the unit circle. The buses have their own intrinsic natural frequencies $\omega_1>\cdots>\omega_N$ (excluding time stopped at the bus stops). If the buses have two doors: one for passengers to alight and the other to board with alighting and boarding occuring simultaneously, then there is a critical transition defined by $k=k_c$ where the buses would exhibit \emph{complete synchronisation} if $k>k_c$. Here, $k:=s/l$ is a parameter that determines the strength of coupling amongst the $N$ buses due to the $M$ bus stops, with $s$ being the rate of passengers arriving at each of the $M$ bus stops and $l$ being the rate of loading/unloading of people onto/off the bus. Ref.\ \cite{Vee2019} has calculated this critical transition to be given by:
\begin{align}\label{synkc}
k_c(N)=\frac{1}{M}\sum_{i=1}^{N-1}\left(1-\frac{\omega_N}{\omega_i}\right).
\end{align}
Here, $\omega_i=2\pi f_i=2\pi/T_i$, for each $i=1,\cdots,N$.

If $k>k_c$, then all $N$ buses are synchronised and phase-locked with phase difference $\sim0^\circ$. By numerical simulations, Ref.\ \cite{Vee2019} further found that if $\bar{k}<k<k_c(N)$ where $\bar{k}\approx k_c(2)$, with $k_c(2)$ being the corresponding system with $N=2$ buses having natural frequencies $\omega_1$ and $\omega_N$, respectively, then these $N$ buses are \emph{partially synchronised} with some buses being phase-locked (phase difference $\sim0^\circ$) whilst the others are able to pull away (opening up large phase differences). If $k<\bar{k}$, then all $N$ buses are \emph{unsynchronised} with no phase-locking. The phase $k<\bar{k}$ (note the dual usage of the word ``phase'') is referred to as \emph{lull} since $k$ is weak enough such that the demand for service is low and the buses do not persistently bunch (no synchronisation). If $k>\bar{k}$ (i.e. both $\bar{k}<k<k_c$ and $k>k_c$), then this phase is referred to as \emph{busy} since demand for service is high enough leading to the undesirable phenomenon of \emph{persistent} bus bunching of at least one pair of buses due to phase-locking. Intriguingly, Ref.\ \cite{Vee2019} empirically observed \emph{real human-driven buses} serving $M=12$ reasonably staggered bus stops in a loop service being: (a) in the lull phase with measured $k_\textrm{measured}<\bar{k}$, where those buses are observed to be not exhibiting persistent bunching; (b) in the busy phase with measured $k_\textrm{measured}>\bar{k}$, where some of those buses are indeed observed to be phase-locked and persistently bunched.

Incidentally, note that if boarding and alighting occur sequentially through one single door, i.e. people alight first and then followed by people boarding, there will be an overall factor of $1/2$ in Eq.\ (\ref{synkc}) as can be verified by direct calculations analogous to that carried out in Ref.\ \cite{Vee2019} [or see the description below Eq.\ (\ref{NBkc})], since only half of the time stopped at the bus stop is devoted to boarding the people waiting there.

\section{Fourier expansion for the bus loop system}

Consider $N=1$ bus serving $M=1$ bus stop in a loop. This bus moves with its constant natural (angular) frequency $\omega$ when it is not at the bus stop, and stops for a duration of $\tau$ to embark/disembark passengers at the bus stop. This repeats over and over as it loops around the circle. Hence, we can write:
\begin{align}\label{Fourierbus}
\frac{d\theta}{dt}=\omega-\omega f(t),
\end{align}
where
\begin{align}
f(t):=\frac{\tau}{T}+\sum_{n=1}^{\infty}{\frac{2}{n\pi}\sin{\left(\frac{n\pi\tau}{T}\right)}\cos{\left(\frac{2n\pi}{T}t\right)}}.
\end{align}
This function $f(t)$ is a pulse wave with period $T=2\pi/\omega$ and pulse time $\tau$.

In general with $N$ buses, each bus's rate of change of $\theta_i$ takes the form of Eq.\ (\ref{Fourierbus}), with the width of the well $\tau_i$ depending on the headway from the bus immediately ahead of it. Therefore, $\tau_i$ is essentially proportional to $\theta_{i-1}-\theta_i$, i.e. the phase difference with respect to the bus immediately ahead of it. The larger the phase difference since the bus immediately ahead has left the bus stop, the more people would have accumulated at the bus stop to be picked up. These buses can overtake, and the identity of the bus immediately ahead must change accordingly. With $M$ bus stops on the loop, there will be $M$ such wells. Thus we find this system to be governed by a complicated coupled set of nonlinear differential equations with feedback that is non-analytic due to the change of identity corresponding to $\theta_{i-1}$ each time overtaking occurs.

\section{Local unidirectional Kuramoto oscillators}

Kuramoto studied a simple model of coupled oscillators where he specified only the first sine term in the Fourier expansion for the coupling function \cite{Kura84}. We would like to investigate the corresponding Kuramoto version for the bus loop system and find out which properties of the bus dynamics correspondingly exist in this simplified model with only the first sine term in the coupling function. Consider $N$ Kuramoto oscillators with natural angular frequencies $\omega_1,\cdots,\omega_N$, subjected to local unidirectional coupling. This system differs from the original Kuramoto model (with the exception of $N=2$ which is indeed equivalent), whereby here oscillator $i$, $i\in\{1,\cdots,N\}$ is only directly influenced by the oscillator that is immediately ahead of it. We are interested in this type of Kuramoto oscillators instead of the original model whereby the coupling is globally contributed by every oscillator, because this is the direct analogue to the bus system (a bus is only directly influenced by the bus immediately ahead).

In accordance to Refs.\ \cite{Syn03,Rogge04}, we say that a pair of adjacent oscillators are \emph{phase-locked} if their phase difference remains constant in time. If the coupling is \emph{strong enough}, i.e. $K\geq K_c$, then these $N$ oscillators are \emph{completely phase-locked}. The critical coupling $K_c$ is the value of $K$ such that the system is barely able to be completely phase-locked. Hence, the $N$ equations governing the motion of the $N$ oscillators in such a situation are:
\begin{align}
\frac{d\theta_i}{dt}=\omega_i+K\sin{(\theta_{i-1}-\theta_i)},
\end{align}
where $\theta_i$ represents the phase of oscillator $i$ on the unit circle, and $\theta_0:=\theta_N$. The constant $K$ is the coupling strength between the oscillators. Note that we do not divide $K$ by $N$ (which is done in the original Kuramoto model), because there is only one sine term present (instead of $N-1$ sine terms in the original Kuramoto model). The condition for these oscillators being completely phase-locked is:
\begin{align}
\frac{d\theta_1}{dt}=\cdots=\frac{d\theta_N}{dt},
\end{align}
giving $N-1$ independent equations which are functions of $N-1$ independent phase differences $\phi_1=\theta_1-\theta_2, \cdots, \phi_{N-1}=\theta_{N-1}-\theta_N$.

Let us work this out explicitly in the case of $N=2$. The governing equations are:
\begin{align}
\frac{d\theta_1}{dt}&=\omega_1+K\sin{(\theta_2-\theta_1)}=\omega_1-K\sin(\theta_1-\theta_2)\\
\frac{d\theta_2}{dt}&=\omega_2+K\sin{(\theta_1-\theta_2)}.
\end{align}
The phase-locked condition gives:
\begin{align}
\frac{d\theta_1}{dt}&=\frac{d\theta_2}{dt}\label{phaselockingcondition}\\
K&=\frac{\omega_1-\omega_2}{2\sin{(\theta_1-\theta_2)}}.\label{KforN2}
\end{align}
This implies that the two oscillators can be phase-locked as long as there is a phase difference $\theta_1-\theta_2$ which satisfies Eq.\ (\ref{KforN2}), given some value of $K$. If $K$ is not strong enough, i.e. $K<K_c$, then no phase difference $\theta_1-\theta_2$ can satisfy Eq.\ (\ref{KforN2}) and the two oscillators are not phase-locked. The sought after critical transition $K_c$ between the unsynchronised and synchronised phases of the system occurs when Eq.\ (\ref{phaselockingcondition}) is satisfied for the smallest $K$, i.e. we need to minimise Eq.\ (\ref{KforN2}). This gives
\begin{align}\label{KcsolutionN2}
K_c=\frac{1}{2}(\omega_1-\omega_2),
\end{align}
which occurs when the phase difference between the two oscillators is $\theta_1-\theta_2=\pi/2$.

For $N=3$, the governing equations are:
\begin{align}
\frac{d\theta_1}{dt}&=\omega_1+K\sin{(\theta_3-\theta_1)}=\omega_1-K\sin(\phi_1+\phi_2)\\
\frac{d\theta_2}{dt}&=\omega_2+K\sin{(\theta_1-\theta_2)}=\omega_2+K\sin{\phi_1}\\
\frac{d\theta_3}{dt}&=\omega_3+K\sin{(\theta_2-\theta_3)}=\omega_3+K\sin{\phi_2},
\end{align}
where $\phi_1=\theta_1-\theta_2$, $\phi_2=\theta_2-\theta_3$, $\theta_3-\theta_1=-(\theta_1-\theta_2+\theta_2-\theta_3)=-(\phi_1+\phi_2)$. From the condition for complete phase-locking
\begin{align}
\frac{d\theta_1}{dt}=\frac{d\theta_2}{dt}=\frac{d\theta_3}{dt},
\end{align}
we have two independent equations
\begin{align}
K&=\frac{\omega_1-\omega_2}{\sin{\phi_1}+\sin{(\phi_1+\phi_2)}}\\
K&=\frac{\omega_2-\omega_3}{\sin{\phi_2}-\sin{\phi_1}}.
\end{align}
This implies that the three oscillators can all be phase-locked if there are some phase differences $\phi_1$ and $\phi_2$ between these three oscillators which satisfy the above two equations, given some value of $K$. The sought after critical transition between completely phase-locked and partially phase-locked phases of the system occurs when $K$ is minimised. Since $K$ depends on the two variables $\phi_1,\phi_2$ which are subjected to the constraint
\begin{align}
\frac{\omega_1-\omega_2}{\sin{\phi_1}+\sin{(\phi_1+\phi_2)}}=\frac{\omega_2-\omega_3}{\sin{\phi_2}-\sin{\phi_1}},
\end{align}
the problem of minimising $K$ can be solved by the method of Lagrange multipliers. Unfortunately, there does not appear to be a closed form solution for $K_c$. This applies to any $N$ oscillators, where $K$ is a function of $N-1$ independent phase differences which are subjected to $N-2$ constraint equations. The closed form solution is found for $N=2$ in Eq.\ (\ref{KcsolutionN2}), but not for $N>2$.

Interestingly for $N=3$, we note that if $\omega_1-\omega_2=\omega_2-\omega_3=\Omega$, then $\phi_1=0$, $\phi_2=\pi/2$ is the solution that minimises $K$ as can be verified by the method of Lagrange multipliers. In this case,
\begin{align}
K_c=\Omega.
\end{align}
However, the corresponding situations with $\omega_1-\omega_2=\cdots=\omega_{N-1}-\omega_N$ for $N>3$ do not seem to admit such a nice closed form solution. Nevertheless, these can be solved numerically as studied by Ref.\ \cite{Rogge04} for the unidirectionally coupled case (like our formulation here where we are motivated by the bus loop system) and Ref.\ \cite{Ochab10} for a locally (nearest neighbour) coupled case.

\section{Partial synchronisation of local unidirectional Kuramoto oscillators}

As mentioned in the Introduction, the bus loop system has three phases: (a) complete phase-locking if $k>k_c(N)$; (b) partial phase-locking if $\bar{k}<k<k_c(N)$; (c) no phase-locking if $k<\bar{k}$; with $\bar{k}\sim k_c(2)$. The value of $\bar{k}\sim k_c(2)$ is a result found from simulations \cite{Vee2019}. Here, we similarly carry out simulations for the local unidirectional Kuramoto model to investigate if partial synchronisation occurs. We consider three situations:
\begin{enumerate}
\item Frequencies skewed towards lowest frequency.
\item Evenly spaced frequencies.
\item Frequencies skewed towards highest frequency.
\end{enumerate}

\begin{figure}
\centering
\includegraphics[width=16cm]{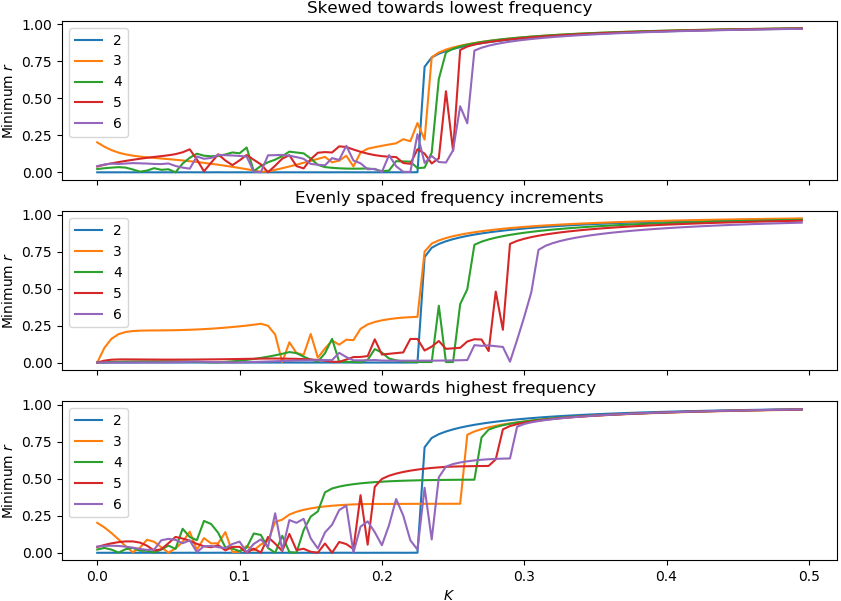}
\caption{Partial synchronisation in the local unidirectional Kuramoto oscillators are observed when the natural frequencies are skewed towards the highest one. Shown here are simulation runs with $N=2,3,4,5,6$ oscillators, respectively. Regions of partially phase-locked configurations with $0\ll r\ll1$ are seen in the bottom plot, whereas partial phase-locking is absent in the first two graphs since $r$ essentially jumps from $\sim0$ to a value near $\sim1$.}
\label{fig1}
\end{figure}

These frequencies are selected with $\omega_1=1.39$, $\omega_N=0.93$, and $\omega_2,\cdots,\omega_{N-1}\in(\omega_1,\omega_N)$. (Note: The frequency range of a university bus loop service studied in Ref.\ \cite{Vee2019} was measured to be $[0.93,1.39]$ mHz, i.e. a period range of $[12,18]$ minutes, excluding time stopped at bus stops.) Simulations results as shown in Fig.\ \ref{fig1} indicate that unlike the bus loop system \cite{Vee2019}, the local unidirectional Kuramoto oscillators only exhibit partial synchronisation if the frequencies are skewed towards the highest frequency. Even so, the corresponding $\bar{K}$ where partial synchronisation starts to occur does not seem to be near $K_c(2)$. The order parameter $r\in[0,1]$ measures the degree of synchronicity amongst the $N$ oscillators. If they are fully synchronised, then $r=1$, whilst if they are fully unsynchronised then $r=0$. Its definition is
\begin{align}
r=\frac{1}{N}\sqrt{\left(\sum_{i=1}^{N}{\cos{\theta_i}}\right)^2+\left(\sum_{i=1}^{N}{\sin{\theta_i}}\right)^2}.
\end{align}
The graphs measure the minimum $r$ in the latter part of the simulations where the transients have weeded out. If there is some degree of phase-locking, then the minimum $r$ should report a value larger than $0$, whilst a minimum $r\sim0$ implies that there is no synchronisation since the oscillators overlap one another as well as open up large phase differences.

\section{Anti-bus bunching}\label{noboarding}

Several ways to overcome the notorious phenomenon of bus bunching have been proposed over the decades (see the introduction and references in Ref.\ \cite{Vee2019}). A recent idea \cite{Vee2019b} explores a no-boarding strategy where a ``slow'' bus, i.e. one whose phase difference from the bus immediately behind it is way too small (or alternatively, if its phase difference from the bus immediately ahead is way too large), would disallow boarding and leave, leaving the remaining people to be picked up by the bus behind it who would be arriving soon. An analytical study in Ref.\ \cite{Vee2019b} showed that this strategy can significantly improve the efficiency of the bus system compared to leaving the buses on their own, in the case where all buses have an identical natural frequency. This is useful for the future when self-driving buses are the norm and can be programmed to maintain an identical natural frequency, unlike human-driven buses which tend to have different natural frequencies due to differing driving styles \cite{Vee2019}. The no-boarding policy also works well for buses with frequency detuning when the demand level $k$ is high, but surprisingly turns out to backfire when $k$ is low. This section establishes the phase transition when this happens. Further work \cite{Vee2019d} showed that self-learning buses without human instruction also end up discovering a no-boarding strategy as well as a holding strategy to maintain a staggered configuration around the loop and prevent bus bunching. On top of that, as a mandatory no-boarding policy is arguably stringent because it denies commuters who need service urgently, we have studied the situations where the commuters are allowed to cooperate or defect the implementation of no-boarding, and this turns out to be a reasonably viable approach \cite{Vee2019c}.

Consider the simplest setup of $N=2$ buses with natural angular frequencies $\omega_1>\omega_2$ ($\omega_i=2\pi f_i=2\pi/T_i$) serving $M=1$ bus stop in a loop. When a bus is at a bus stop, it would allow people to alight followed by allowing people to board, i.e. the two processes are sequential. According to the no-boarding policy, however, whilst boarding people it would disallow further boarding if its phase difference $\Delta\theta$ as measured from the bus behind it is less than a prescribed threshold $\theta_0$. The intention of the no-boarding policy is to speed up this bus which is deemed as being too slow when $\Delta\theta<\theta_0$, leaving the people who are denied boarding to be picked up by the ``faster'' bus which would be approaching soon. Consequently, these two buses would reach a steady state whereby they both maintain a staggered configuration with an effective average phase difference of $\theta_\textrm{eff}>\theta_0$. The effective average phase difference $\theta_\textrm{eff}$ is greater than $\theta_0$ where no-boarding is implemented since $\theta_0$ serves as the lower bound to the evolving phase difference $\Delta\theta(t)$ which fluctuates around $\theta_\textrm{eff}$. In an ideal implementation, the system would be fully synchronised and the $N$ buses are phase-locked, with nearly staggered phase difference of $\theta_\textrm{eff}\sim360^\circ/N$ depending on the prescribed $\theta_0$.

The no-boarding policy always works, regardless of $k$, if the two buses have the same natural frequency $\omega$, provided that the prescribed $\theta_0$ does not exceed an upper bound. This upper bound exists because if $\theta_0$ is too large, then the buses would implement no-boarding too frequently and fail to board people at a rate faster than the rate of people arriving at the bus stop (see details in Ref. \cite{Vee2019b}). In contrast, this is not true if the buses have frequency detuning. The slow bus with $\omega_2$ can only be sped up up to a certain extent, leaving the people denied boarding to be picked up by the fast bus with $\omega_1$ which only slows it down up to a certain extent.

It turns out that there is a critical $k_c$ [which we will determine below, and turns out to be the same $k_c$ in Eq.\ (\ref{synkc}) with a factor of 1/2] whereby if $k<k_c$, then the fast bus is not slowed down enough due to not that many people to pick up (i.e. the coupling strength is too weak). This situation can also be viewed as the frequency detuning being too large for the no-boarding policy to bridge. In this phase, the two buses would fail to maintain their staggered configuration with an effective average phase difference of $\theta_\textrm{eff}$. Instead, the fast bus would periodically overlap the slow bus, with these two buses being unsynchronised and not phase-locked. On the other hand if $k>k_c$, then the two buses are in a phase whereby they are phase-locked with an effective average phase difference of $\theta_\textrm{eff}$. Thus, the critical $k_c$ is defined by the condition that the slow bus has \emph{zero} stoppage duration at the bus stop since this is the best possible way to speed it up by not having to stop at the bus stop at all. In this extreme no-boarding policy where the slow bus always implements it, all passengers are picked up by the fast bus.


\subsection{Direct calculation}

\begin{figure}
\centering
\includegraphics[width=16cm]{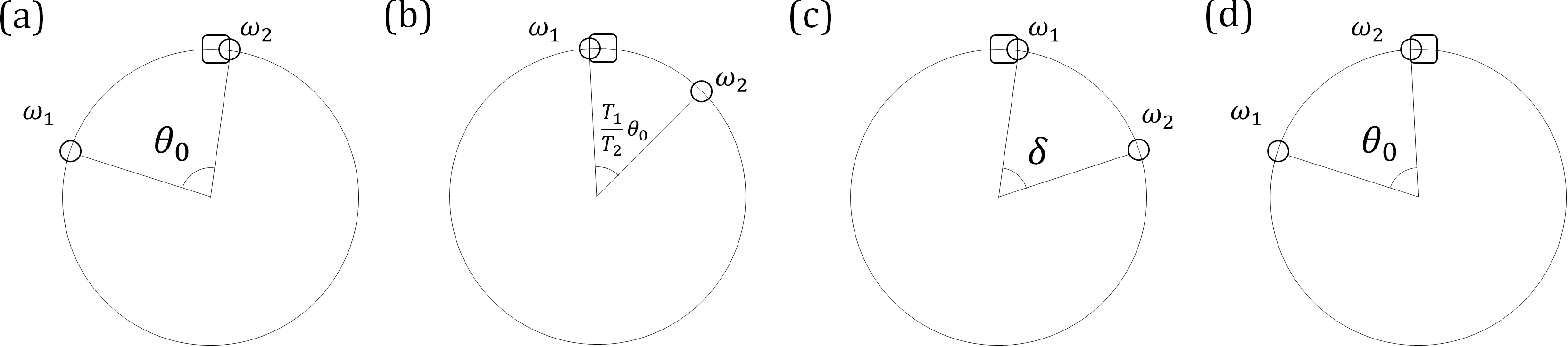}
\caption{$N=2$ buses serving $M=1$ bus stop in a loop, where $k$ is strong enough such that they are phase-locked.}
\label{fig2}
\end{figure}

Let us directly calculate $k_c$. Fig. \ref{fig2} shows a cycle where the two buses are in steady state: (a) is the moment when the slow bus just leaves the bus stop after implementing the no-boarding policy as the phase difference from the bus behind it is $\Delta\theta=\theta_0$, (b) is the moment when the fast bus just arrives at the bus stop, (c) is the moment when the fast bus leaves the bus stop after allowing everyone to alight followed by everyone to board, and (d) is when the slow bus just arrives at the bus stop.

The time elapsed from (a) to (b) is $\theta_0/\omega_1$, so in (b) the phase difference of the slow bus as measured from the bus behind it is $T_1\theta_0/T_2$. The time elapsed from (b) to (c) is $\tau_1$, the stoppage duration that the fast bus spends at the bus stop. This stoppage duration $\tau_1$ at the critical $k_c$ can be calculated as follows: the total number of people arriving at the bus stop per cycle is $s(T_1+\tau_1)$, since one cycle is the natural period of the fast bus, $T_1$, plus the time it stops at the bus stop, $\tau_1$. Half of $\tau_1$ is allocated for people to alight, with the other half for people to board. Hence $l\tau_1/2=s(T_1+\tau_1)$, giving $\tau_1=2k_cT_1/(1-2k_c)$ with $k_c:=s/l$. Consequently in (c), the phase difference of the slow bus as measured from the bus behind it is $\delta=T_1\theta_0/T_2+\omega_2\tau_1$. From (c) to (d), the slow bus has to travel a phase difference of $2\pi-\delta$ which takes a time of $(2\pi-\delta)/\omega_2$, ending up with a phase difference of $\delta-(\omega_1-\omega_2)(2\pi-\delta)/\omega_2$ as measured from the bus behind it. At the critical $k_c$, the slow bus barely arrives at the bus stop and immediately implements no-boarding (note that in the previous round it picked up nobody and therefore has nobody to alight) since its phase difference from the bus behind it is $\theta_0$. This gives the condition for obtaining $k_c$:
\begin{align}\label{deltacondition}
\delta-\frac{(\omega_1-\omega_2)(2\pi-\delta)}{\omega_2}=\theta_0.
\end{align}
Plugging in $\delta=T_1\theta_0/T_2+\omega_2\tau_1$ and $\tau_1=2k_cT_1/(1-2k_c)$, we find that the terms involving $\theta_0$ cancel out, giving:
\begin{align}\label{kc2buses}
k_c=\frac{1}{2}\left(1-\frac{\omega_2}{\omega_1}\right).
\end{align}

\subsection{Alternative condition}

The direct calculation shows that regardless of the angle $\theta_0$ where no-boarding is implemented, the resulting $k_c$ is independent of $\theta_0$. In other words, the two buses can remain phase-locked due to the no-boarding policy at any effective average phase difference $\theta_\textrm{eff}>\theta_0$ as long as $k>k_c$. There is a simpler alternative condition to Eq.\ (\ref{deltacondition}). Since the critical transition is given by the condition whereby the slow bus has zero stoppage duration at the bus stop, thus in steady state, the natural period of the slow bus $T_2$ must equal the natural period of the fast bus $T_1$ plus the stoppage duration of the fast bus $\tau_1$:
\begin{align}\label{simplercondition}
T_2=T_1+\tau_1,
\end{align}
where $\tau_1=2k_cT_1/(1-2k_c)$. The alternative condition Eq.\ (\ref{simplercondition}) also leads to the same formula for $k_c$ in Eq.\ (\ref{kc2buses}), with no dependence on $\theta_0$.

In fact, this approach is directly generalisable to any $N$ buses with natural frequencies $\omega_1>\cdots>\omega_N$ serving one bus stop in a loop. If $k>k_c$, then all buses can remain staggered and phase-locked with phase difference $\theta_\textrm{eff}>\theta_0$ (for any $\theta_0$) due to the no-boarding policy. At the critical transition $k=k_c$, the slowest bus is sped up in the best possible way whereby its stoppage duration is zero. Therefore, we have $N-1$ (independent) equations when the $N$ buses are in steady state:
\begin{align}\label{N-1simplerconditions}
T_N=T_i+\tau_i,
\end{align}
where $i=1,2,\cdots,N-1$, and $\tau_i$ are the stoppage durations for the other buses with natural frequencies $\omega_1>\cdots>\omega_{N-1}$, respectively. These $\tau_i$ must satisfy:
\begin{align}\label{totaltaus}
\frac{1}{2}l\sum_{i=1}^{N-1}{\tau_i}=sT_N,
\end{align}
since half of all these stoppage durations would be used to board the people that arrive at the bus stop during one cycle of $T_N$ (with the other half meant for alighting). Plugging in the $N-1$ equations for each $\tau_i$ from Eq.\ (\ref{N-1simplerconditions}) into Eq.\ (\ref{totaltaus}):
\begin{align}
\frac{1}{2}\sum_{i=1}^{N-1}{(T_N-T_i)}=k_cT_N,
\end{align}
where $k_c:=s/l$, gives the general formula for $k_c$:
\begin{align}
k_c=\frac{1}{2}\sum_{i=1}^{N-1}{\left(1-\frac{\omega_N}{\omega_i}\right)}.
\end{align}
With $M$ staggered bus stops in the loop each having a spawning rate of $s$, then each bus stop acts as a multiplier of the coupling strength. Hence, only one $M$th of the coupling strength is required for the critical transition, i.e.
\begin{align}\label{NBkc}
k_c=\frac{1}{2M}\sum_{i=1}^{N-1}{\left(1-\frac{\omega_N}{\omega_i}\right)},
\end{align}
where $\omega_N/\omega_i=f_N/f_i=T_i/T_N$.

Remarkably, this is the identical formula as Eq. (\ref{synkc}) with the overall factor of $1/2$ since alighting and boarding here occur sequentially via one door instead of simultaneously via two doors. This coincidence is due to the fact that at the critical transition to complete synchronisation, the coupling strength is just enough to keep all $N$ buses bunched together at the same bus stop. (See Ref.\ \cite{Vee2019} for the original derivation which directly goes through the process for one cycle.) Therefore, the slowest bus would experience \emph{zero stoppage duration} in steady state, i.e. we would arrive at the $N-1$ (independent) equations given by Eq.\ (\ref{N-1simplerconditions}), as well as all the stoppage durations $\tau_i$ ($i=1,\cdots,N-1$) for the other $N-1$ buses satisfying Eq.\ (\ref{totaltaus}). This would lead to the same formula for $k_c$ in Eq.\ (\ref{NBkc}) for $N$ buses serving $M$ bus stops.

\section{Stable staggered solutions of the local unidirectional Kuramoto oscillators}

Work by Ref.\ \cite{Rogge04} studied the stability of completely phase-locked unidirectional Kuramoto oscillators. Their results reveal the intriguing implications on the anti-bunched or staggered configuration of the oscillators:

\vskip 0.5 cm

\emph{A completely phase-locked solution with all phase differences between adjacent oscillators being less than $\pi/2$ is asymptotically stable.}

\vskip 0.5 cm

To illustrate, consider a system of $N=3$ Kuramoto oscillators with identical natural frequency $\omega$. If these three oscillators are such that they are all $120^\circ$ spaced apart, then they are completely phase-locked, i.e. they will always remain staggered. However, if they are slightly perturbed, then they will deviate away from this staggered configuration and end up in a fully synchronised configuration with $r=1$ or $\theta_1=\cdots=\theta_N$. Intriguingly if $N\geq5$, then the staggered configuration is \emph{stable} since all phase differences between adjacent oscillators are strictly less than $\pi/2$. If the system is slightly perturbed, then it will return to this staggered configuration. For more details and other stable/unstable conditions, see Ref.\ \cite{Rogge04}.

\begin{figure}
\centering
\includegraphics[width=10cm]{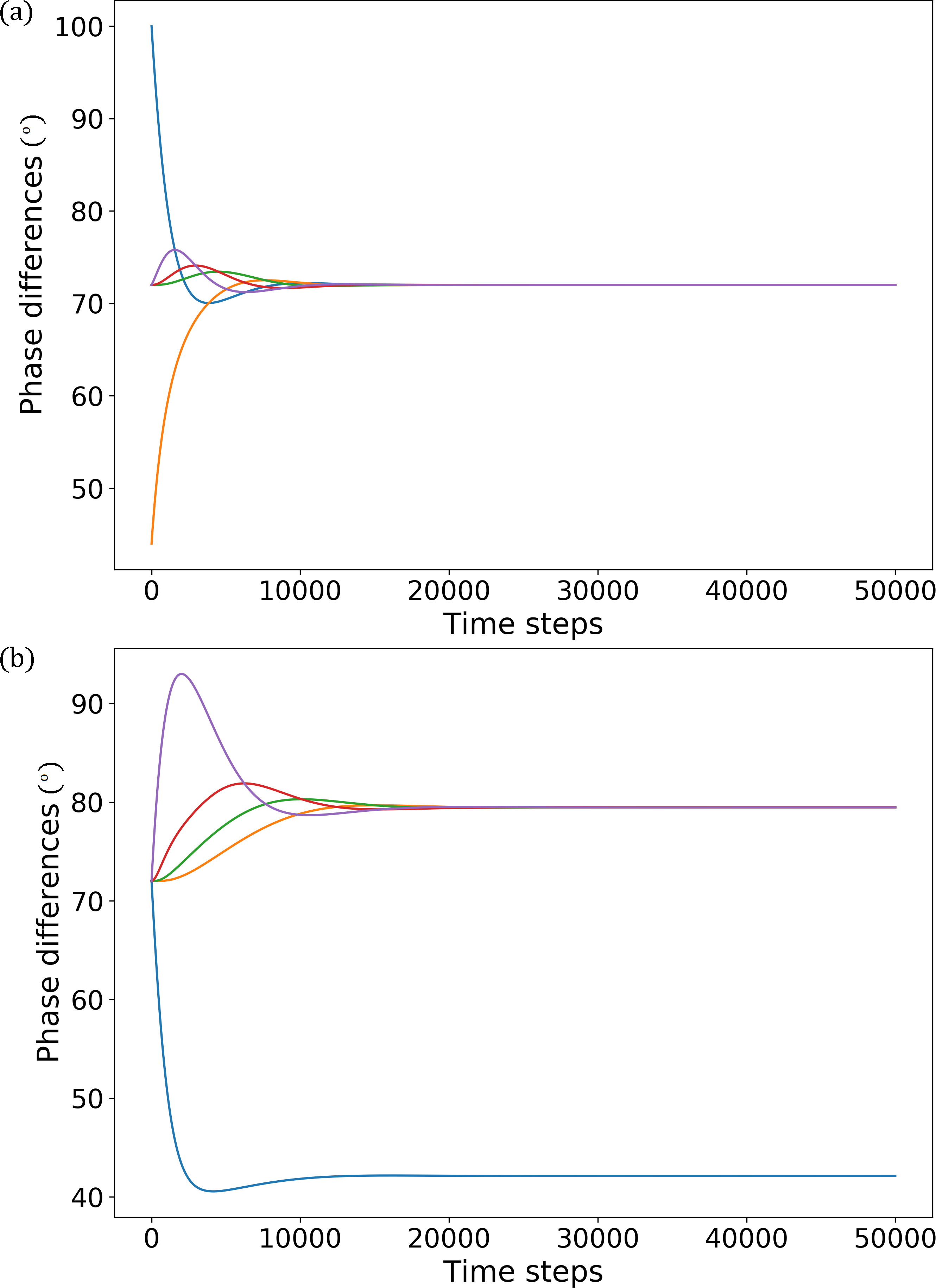}
\caption{Graphs of the phase differences for $N=5$ local unidirectionally coupled Kuramoto oscillators. In both configurations, they remain anti-bunched with all phase differences less than $\pi/2$.}
\label{fig3}
\end{figure}

Fig.\ \ref{fig3} (videos are given as supplemental material) shows the situations with $N=5$ oscillators and $K=0.32$ where: (a) They have identical natural frequency of $\omega=1$, and their initial positions are $0^\circ,100^\circ,144^\circ,216^\circ,288^\circ$; and (b) Four oscillators have identical natural frequency $\omega=1$ with the other having $\omega_\textrm{other}=1.1$, and their initial positions are perfectly staggered. In both cases, they remain anti-bunched with all phase differences less than $\pi/2$ --- in accordance to the theorem by Ref.\ \cite{Rogge04}. Simulations with these setups for the original globally coupled Kuramoto oscillators \cite{Kura84} reveal that they would all end up bunching with $r\sim1$.

The intuition of this stability property can be grasped from the nature of the sine coupling function [see also Fig.\ \ref{fig4}(a)]: If the phase difference is less than $\pi/2$, the trailing oscillator is given a positive kick to catch up. The larger the phase difference, the stronger is the kick due to the coupling it receives to speed it up. However, $\pi/2$ is a \emph{turning point}. If the phase difference exceeds $\pi/2$, it lags even more but then receives \emph{weaker kick} due to the coupling --- which aggravates its lagginess. This is why the system with $N=3$ staggered oscillators with identical natural frequency $\omega$ is unstable, because if one oscillator experiences a small lag, then the coupling it receives becomes weaker and it lags even more until they all crash into the fully synchronised configuration.

Ref.\ \cite{Rogge04} further extended their results to a broader class of coupling functions which are continuous, $2\pi$-periodic, odd, and has exactly one maximum at  $\gamma\in(0,\pi)$. Then, a completely phase-locked solution with all phase differences between adjacent oscillators being less than $\gamma$ is asymptotically stable. So instead of receiving the maximum boost when the phase difference is $\pi/2$, here it occurs at $\gamma$.

\section{Creating stability for a staggered configuration in the bus loop system}

Eq.\ (\ref{Fourierbus}) describes the instantaneous angular velocity experienced by the bus in terms of the pulse waveform where the bus moves with its constant natural frequency when it is not at a bus stop, and has zero angular velocity when it is at a bus stop over a dwell time of $\tau$ which depends on the headway (phase difference) from the bus immediately ahead of it.

A different view for the bus loop system may be seen from considering the average angular velocity over a time interval. For example, suppose there are $N$ buses serving $M$ staggered bus stops in a loop where each bus stop has people arriving at a rate of $s$ people per second. When a bus just leaves a bus stop, suppose it takes a time of $\Delta t$ before the next bus arrives at the bus stop and then leaves after disembarkation/embarkation. The number of people that this bus picks up is $s\Delta t$. If this bus spends a dwell time of $\tau$, then $l\tau\sim2s\Delta t$ where $l$ is the rate of loading/unloading. The factor of $2$ on the right hand side is because one factor is expended to pick up $s\Delta t$ people with the other for those alighting. With $k:=s/l$, we have $\tau\sim2k\Delta t$. Since its natural frequency is $\omega_i$, its natural period is $2\pi/\omega_i$. As there are $M$ staggered bus stops in the loop, the time it takes to traverse one bus stop would be $2\pi/M\omega_i+\tau$, and it traverses an angular displacement of $2\pi/M$. Let $\eta_i=2\pi/M\omega_i$, i.e. the time interval to traverse one bus stop excluding the dwell time $\tau$. Thus, the average angular velocity is:
\begin{align}
\left\langle\frac{d\theta_i}{dt}\right\rangle&=\frac{\omega_i}{1+2k\Delta t/\eta_i}\\
&=\omega_i\left(1-\frac{1}{1+\eta_i/2k\Delta t}\right).\label{averageomega}
\end{align}
We have in fact derived the result given by Eq.\ (2) of Ref.\ \cite{Vee2019}. (Recall that in Ref.\ \cite{Vee2019}, people alight/board via two different doors simultaneously whereas here we have people first alighting and then boarding, sequentially, via one door. So here we have a factor of $2$ in the expression for $\tau$.) In this expression for the average angular velocity $\langle d\theta_i/d t\rangle$, the second term on the right hand side $-\omega_i/(1+\eta_i/2k\Delta t)$ is the coupling amongst the buses due to the bus stop, which is directly analogous to the sine coupling of the Kuramoto oscillators.

\begin{figure}
\centering
\includegraphics[width=16cm]{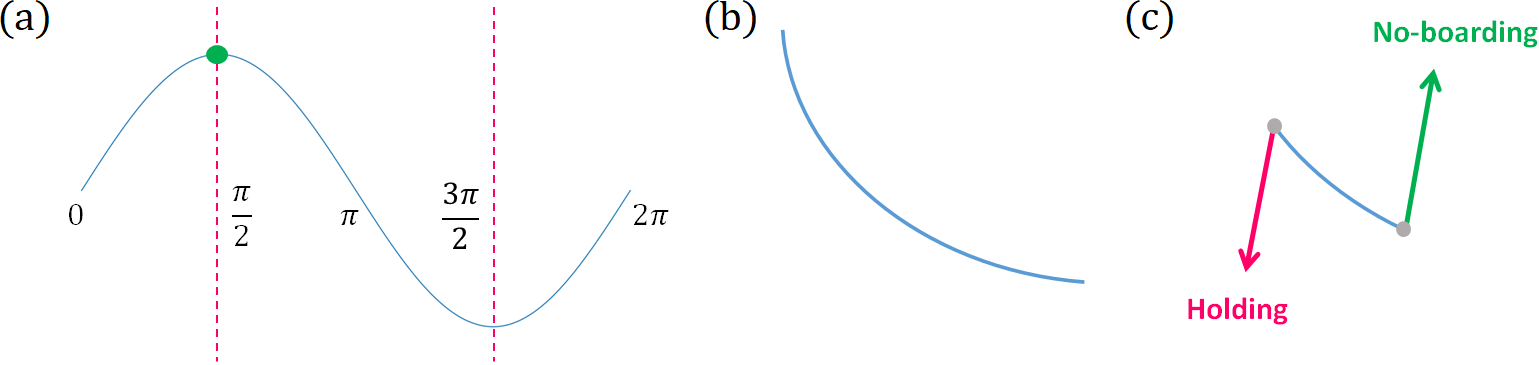}
\caption{(a) The local unidirectional Kuramoto oscillators experience a sine coupling with a local maximum at $\Delta\theta=\pi/2$. (b) The average angular velocity for the buses is monotonically decreasing with $\Delta\theta$. (c) The no-boarding strategy creates stability of the anti-bunched configuration by modifying this curve such that a slow bus gets a stronger kick to speed it up. The holding strategy requires a fast bus to slow down, preventing imminent bunching. Overall, the curve becomes sine-like.}
\label{fig4}
\end{figure}

The key point about Eq.\ (\ref{averageomega}) is that the average angular velocity to traverse a bus stop is a function which \emph{monotonically decreases} with the time headway $\Delta t$ for any $\Delta t>0$, where the time headway is essentially the phase difference $\Delta\theta$ with respect to the bus immediately ahead of it [Fig.\ \ref{fig4}(b)]. In comparison with the Kuramoto oscillators, here the expression for $\langle d\theta_i/d t\rangle$ has no local maximum within $\Delta\theta\in(0,\pi)$ (the maximum at $\Delta\theta=0$ does not count). Any perturbed lag on a trailing bus (i.e. a slight increase in $\Delta t$ or $\Delta\theta$) would necessarily cause it to move slower, further lagging behind (i.e. opening up larger $\Delta t$ or $\Delta\theta$), and move even slower $\cdots$ until they bunch. This explains why the bus system cannot have a stable staggered configuration, leading to the perennial and notorious phenomenon of bus bunching which plagues bus systems worldwide.

The no-boarding strategy modifies the monotonically decreasing $\langle d\theta_i/d t\rangle$ with respect to $\Delta t$ (or $\Delta\theta$) in Eq.\ (\ref{averageomega}) by speeding up a bus when it is too slow [Fig.\ \ref{fig4}(c)]. For instance, if its phase difference with respect to the bus immediately ahead of it gets too large, then it activates the no-boarding strategy to speed it up --- in other words if $\Delta\theta$ grows then the coupling \emph{increases} $\langle d\theta_i/d t\rangle$.

Note however, that the stability results from Ref.\ \cite{Rogge04} only applies if there is such a completely phase-locked solution in the first place. Our calculations in Section\ \ref{noboarding} established that there is a critical $k_c$ given by Eq.\ (\ref{NBkc}) such that the no-boarding strategy only works to create such an anti-bunched or staggered configuration if $k>k_c$. If $k$ is too small or the frequency detuning is too great such that $k<k_c$, then there is no such anti-bunched solution, and according to Eq.\ (\ref{synkc}), this is the lull period where the buses are completely unsynchronised and periodically lap one another. This lull period $k<k_c$ is thus a phase where it appears to defy the possibility for attaining the anti-bunched configuration.

Intriguingly, an approach where buses are allowed to learn on their own by reinforcement learning and without any human instruction turns out to discover novel solutions comprising the \emph{no-boarding strategy} as well as a \emph{holding strategy} which allows for stable anti-bunched configurations even in the lull period where $k<k_c$ \cite{Vee2019d}! The holding strategy is implemented by a bus when it is too fast. The way it works in the lull period where the buses have too much frequency detuning is that a fast bus with higher natural frequency slows itself down to reduce the effective frequency detuning of the buses --- which, by Eq.\ (\ref{NBkc}) would reduce $k_c$, and therefore \emph{allows for the existence of anti-bunched solutions}. The bus system is then able to maintain a staggered configuration. Also, holding makes a bus reduce its speed, when its phase difference from the bus ahead of it is too small. Thus overall, the holding and no-boarding strategies make the shape of the curve given by Eq.\ (\ref{averageomega}) to be sine-like instead of monotonically decreasing [Fig.\ \ref{fig4}(c)].

In conclusion, the no-boarding together with the holding strategies, when adaptively and appropriately applied in various situations, are able to allow for the buses to overcome the problem of bus bunching. The correspondence with the local unidirectional Kuramoto oscillators and their stability properties \cite{Rogge04} has led us to understand why a typical bus system always has the bus bunching problem, and let us draw insightful physical understanding on how the bus system can achieve a stable anti-bunched solution. In particular, it provides an explanation to why the simulated buses arrive at the no-boarding and holding strategies by self-reinforcement learning \cite{Vee2019d}, since these two strategies create stable anti-bunched configurations by modifying the shape of the effective average angular velocity to be sine-like --- like that of the local unidirectionally coupled Kuramoto oscillators.



\section*{Supplemental material}

Videos on the $N=5$ local unidirectionally coupled Kuramoto oscillators with stable anti-bunched configurations are found here: \url{https://www.youtube.com/playlist?list=PLZIj25fISvwOALT0PADz_LiH0YbgKMkBz}

\bibliographystyle{naturemag}

\bibliography{Citation}

\begin{acknowledgments}
This work was supported by the Joint WASP/NTU Programme (Project No. M4082189) and the DSAIR@NTU Grant (Project No. M4082418).
\end{acknowledgments}

\end{document}